\documentclass{elsart}
\usepackage{epsfig}
\usepackage{rotating}
\usepackage{lineno}
\begin{document}
\runauthor{Marco Battaglia}
\begin{frontmatter}
\title{Characterisation of\\ a CMOS Active Pixel Sensor\\ for use in the TEAM Microscope}
\author[LBNL,UCSC]{Marco Battaglia\corauthref{cor}},
\corauth[cor]{Corresponding author, Address: Lawrence Berkeley National Laboratory, 
Berkeley, CA 94720, USA, Telephone: +1 510 486 7029.} 
\ead{MBattaglia@lbl.gov}
\author[LBNL]{Devis Contarato},
\author[LBNL]{Peter Denes},
\author[LBNL]{Dionisio Doering},
\author[LBNL]{Thomas Duden},
\author[LBNL]{Brad Krieger},
\author[LBNL,Padova]{Piero Giubilato},
\author[LBNL]{Dario Gnani},
\author[LBNL]{Velimir Radmilovic}
\address[LBNL]{Lawrence Berkeley National Laboratory, 
Berkeley, CA 94720, USA}
\address[UCSC]{Santa Cruz Institute of Particle Physics, University of California at 
Santa Cruz, CA 95064, USA}
\address[Padova]{Dipartimento di Fisica, Universit\`a degli Studi, 
Padova, I-35131, Italy}
\begin{abstract}
A 1M- and a 4M-pixel monolithic CMOS active pixel sensor with 9.5$\times$9.5~$\mu$m$^2$ 
pixels have been developed for direct imaging in transmission electron microscopy as 
part of the TEAM project. We present the design and a full characterisation 
of the detector. Data collected with electron beams at various energies 
of interest in electron microscopy are used to determine the detector response. 
Data are compared to predictions of simulation. The line spread function 
measured with 80~keV and 300~keV electrons is (12.1 $\pm$ 0.7)~$\mu$m and 
(7.4 $\pm$ 0.6)~$\mu$m, respectively, in good agreement with our simulation. 
We measure the detection quantum efficiency to be 0.78$\pm$0.04 at 80~keV and 
0.74$\pm$0.03 at 300~keV.
Using  a new imaging technique, based on single electron reconstruction, the line 
spread function for 80~keV and 300~keV electrons becomes (6.7 $\pm$ 0.3)~$\mu$m 
and (2.4 $\pm$ 0.2)~$\mu$m, respectively. The radiation tolerance of the 
pixels has been tested up to 5~Mrad and the detector is still functional 
with a decrease of dynamic range by $\simeq$~30~\%, corresponding to a reduction 
in full-well depth from $\sim$39 to $\sim$27 primary 300 keV electrons, due to leakage 
current increase, but identical line spread function performance.
\end{abstract}
\end{frontmatter}

\typeout{SET RUN AUTHOR to \@runauthor}


\section{Introduction}

The TEAM (Transmission Electron Aberration-corrected Microscope) 
Project~\cite{team} has developed the most powerful electron microscope to
date, with unprecedented sensitivity and spatial resolution,  
$\simeq$~50~pm~\cite{team2} in both TEM (Transmission Electron Microscopy) and 
Scanning TEM (STEM). This was accomplished through advances in electron optics, 
particularly aberration correction, and the TEAM project included the development
of new kinds of specimen stages and detectors.

One of the goals of the TEAM project is to be able to observe the dynamics of 
processes at the atomic scale~\cite{nature,science}, which requires advances over 
conventional TEM detectors. These include film, image plates and phosphors 
fiber-coupled to CCDs.  Each of these techniques has fundamental limitations for 
high-speed in-situ imaging. Film and image plates directly detect electrons with 
high spatial granularity but require relatively long exposures and are obviously 
not high-speed.  Phosphors fiber-coupled to CCDs have a modest time granularity but 
are limited in their point spread function (PSF) and detection quantum efficiency (DQE) 
due to the physics limitations of the steps needed to convert the primary electrons into
charge on the CCD cell. A high-speed detector able to directly detect electrons 
appears an ideal solution for TEM imaging.   
The key requirements for TEAM are a minimum of 1k~$\times$~1k pixels, in order to obtain 
a field of view of $\simeq$~200~Angstrom with a magnified pixel size of 
$\simeq$~3~pixels/0.5~Angstrom (provided the PSF is $\le$1~pixel) and a readout time of 
$\le$~10~ms, providing 100 or more frames/s. As TEAM is designed for material science 
applications, suitable radiation hardness is essential. 
Considering the standard way of operation of experiments in TEM, a radiation tolerance 
of $\ge$~1~Mrad would enable its use for approximately one year, which appears to be a 
valid requirement. 

There has been significant interest in the application of both 
CMOS~\cite{emicro,deptuch,Denes:2007} and hybrid~\cite{fan,faruqi,faruqi2} pixel 
sensors for direct imaging in TEM, to replace conventional CCDs optically-coupled 
to phosphor plates~\cite{review}.
In two recent papers~\cite{Battaglia:2008yt,Battaglia:2009aa} in 
this journal we discussed the design of a CMOS sensor prototype 
with a radiation-hard pixel cell and characterised its response in terms 
of energy deposition and line spread function. We found that 
by proper design of the pixel cell the sensor can be made enough 
radiation tolerant to operate properly up to several Mrad of 
ionising dose and the line spread function for 10~$\mu$m pixel 
varies from 12~$\mu$m to 8~$\mu$m for electrons in the energy 
range 80~keV $< E_{e} <$ 300~keV of interest in TEM. These 
results motivated the development of a larger size pixel chip 
to perform tests of fast imaging at TEAM.
The final TEAM detector contains 2k $\times$ 2k pixels with 
100 frames/s readout. As an intermediate stage, 1k $\times$ 1k 
detectors were fabricated.
In this paper we present the design of the TEAM1k pixel chip
and discuss the results of its characterisation on a TEAM 
project test column.

\section{The TEAM Pixel Sensors}

The TEAM1k and TEAM2k detectors are fabricated in a 0.35~$\mu$m CMOS  
process.  For the TEAM1k, four 1~$\times$~1~cm$^2$ chips are placed within the 
2~$\times$~2~cm$^2$ reticle area and for the TEAM2k the entire reticle is used.  
All chips employ an identical 9.5~$\times$~9.5~$\mu$m$^2$ pixel.  
In the TEAM1k reticle, two of the chips have a 
1,024$\times$1,024 pixel imaging area (``imaging'' version), and the other two chips 
have the same imaging area, except that a central 500~$\mu$m circular area is replaced 
with a simple diode (``diffraction'' version). One ``imaging'' and one ``diffraction'' 
chip are designed with analog and digital sections very similar to the design previously 
reported, as backup designs.  The other two ``imaging'' and ``diffraction'' chips 
have several improvements over previous designs, and these new versions are the ones discussed 
in this paper. The TEAM1k chips are organised as 16 identical slices each with 1,024 rows and 64 
columns. Each slice has a novel analog output buffer, capable of driving a capacitive load of 
up to 20~pF.  
In addition, each slice has a variable bottom-of-column current load, which can ``look ahead'' 
a programmable number of columns. In a conventional active pixel sensors (APS), 
there is a fixed current bias at the 
bottom of the column.  In order to reduce power, as the sensor has to be operated in vacuum,
the standing current in this bias should be small. For speed, though, this current needs to be 
large enough for the signal to slew within the time needed to digitise a pixel. In this implementation, 
a larger bias current is switched on when the given column is selected ($N=0$), and $m$ 
clock cycles before ($N=m$) and then maintained for the 
next clock cycle when the column is selected.
Sending the output to digitising electronics outside of the microscope vacuum requires an 
amplifier capable of providing high currents as needed to drive the capacitive load.  
Again, with the desire 
to minimise power, the output stage consists of two parts: a conventional low-power operational 
amplifier, along with a higher-power slew rate enhancer (SRE). The small-signal amplifier has the 
necessary gain, bandwidth, and noise performance to settle to the required precision, while the SRE 
minimises the fraction of the sample period that is spent slewing a large voltage step. The SRE is 
activated when the error voltage at the small-signal amplifier input exceeds a designed value.

Since charge generation is confined primarily to the thin epitaxial layer, just $\simeq$~14~$\mu$m 
thick, it is possible to remove most of the underlying detector bulk silicon using a 
back-thinning process. The charge collection, noise and charge-to-voltage calibration 
have been studied by characterising a batch of CMOS pixel sensors before and after 
back-thinning and no significant effects have been observed~\cite{Battaglia:2006tf}.
Back-thinning is performed by Aptek Industries~\cite{aptek-ref} using a proprietary technique.
The process has been extensively tested and yields in excess of 90~\% have been obtained for 
thicknesses down to 50~$\mu$m~\cite{Battaglia:2006tf}. TEAM1k detectors are thinned to 50~$\mu$m.
The use of thin sensors minimises the charge spread due to back-scattering in the bulk, thus 
maximising the contrast ratio in TEM imaging.

\section{Response Simulation}

The energy deposition in the sensor active layer and the lateral charge spread 
are simulated with the {\tt Geant4} program~\cite{Agostinelli:2002hh},
using the low-energy electromagnetic physics models~\cite{Chauvie:2001fh}.
The CMOS pixel sensor is modelled according to the detailed geometric structure 
of oxide, metal interconnect and silicon layers, as specified by the foundry.
Electrons are incident perpendicular to the detector plane and tracked through the 
sensor. For each interaction within the epitaxial layer, the energy released and the 
position are recorded. 

Charge collection is simulated with {\tt PixelSim}, a dedicated digitisation 
module~\cite{Battaglia:2007eu}, developed in the {\tt Marlin} C++ reconstruction 
framework~\cite{Gaede:2006pj} as discussed in~\cite{Battaglia:2008yt}.
The simulation has a single free parameter, the diffusion constant 
$\sigma_{{\mathrm{diff}}}$, used to determine the width of the charge carrier cloud. 
Its value is extracted  from data by a $\chi^2$ fit to the pixel multiplicity in 
the clusters obtained for 300~keV electrons. We find $\sigma_{{\mathrm{diff}}}$ = 
(14.5$^{+2.0}_{-1.0}$)~$\mu$m, which is compatible with both the value obtained for 
1.5~GeV $e^-$s with a prototype having 20$\times$20~$\mu$m$^2$ pixels produced in the 
same CMOS process~\cite{Battaglia:2009aa} and that inferred from the diffusion length 
(estimated from the doping in the epitaxial layer) and the measured charge 
collection time~\cite{Battaglia:2008yt}.

\section{Sensor Tests}

The sensor response to ionising radiation is studied with 5.9~keV X-rays and  
electrons of energy ranging from 80~keV to 300~keV.
The microscope tests are carried out in the FEI Titan test column at the National 
Center for Electron Microscopy (NCEM). 
The prototype chip is mounted on a proximity board which is held on the film insertion plate 
of the microscope. The insertion plate is built on a removable mechanical assembly which allows 
easy and quick replacement of the sensor. The plate can be inserted in and retracted from the 
column by means of a pneumatic actuator.

The temperature is monitored during operation by measuring the resistance of a 
temperature-sensitive resistor included in the chip.
In absence of active cooling, the thick sensor operates at a temperature of 
$\simeq$~90~$^{\circ}$C, while the thin sensor between 40 and 50~$^{\circ}$C, depending on 
running conditions. Data are taken for 6.25 and 25~MHz clock frequency and different beam 
intensities. For response characterisation, four contiguous sectors of the chip are tested at 
the same time, the four analog outputs being read out in parallel. Two flex cables, exiting from 
the column through a vacuum feed-through, provide the interconnection with the data acquisition 
and monitoring electronics located outside the microscope. An intermediate board provides the 
chip with the biases and driving clocks, performs signal amplification and deploys differential 
stages for adapting the analog output signals to the data acquisition (DAQ) system. 
 
\subsection{Data Acquisition and Analysis}

For this study a custom DAQ system~\cite{Battaglia:2009zz} is 
used. Data from four of the chip analog outputs are digitised by four 100~MS/s, 14-bit ADCs 
on a custom-designed board interconnected with a commercial development board~\cite{avnet}, 
implementing a Xilinx Virtex-5 FPGA device, 64~Mb of on-board DDRAM memory and communication 
devices. Clock pattern generation and chip control is performed via high-speed LVDS lines by 
the Virtex-5, which also controls and synchronises the acquisition with the ADCs reference clock.

The DAQ system is connected to a control PC via a USB~2.0 bus, for acquisition control and DAQ 
steering through dedicated registers and for data retrieval. The bandwidth for data transfer 
is $\sim$~40 Mbytes/s. 
A graphical user interface based on a series of dedicated C/C++ classes interfaced to 
the ROOT~\cite{Brun:1997pa} framework classes provides an intuitive and easy-to-use system for 
acquisition monitoring, fast data display and on-line processing~\cite{Battaglia:2009zz}. 
Raw data is stored in ROOT format, and further converted in the LCIO~\cite{Gaede:2005zz} format 
for offline analysis.  This is performed with a set of dedicated processors, developed in the 
{\tt Marlin} reconstruction framework. 

\subsection{Noise, Uniformity and Calibration}

\begin{figure}[ht!]
\begin{center}
\epsfig{file=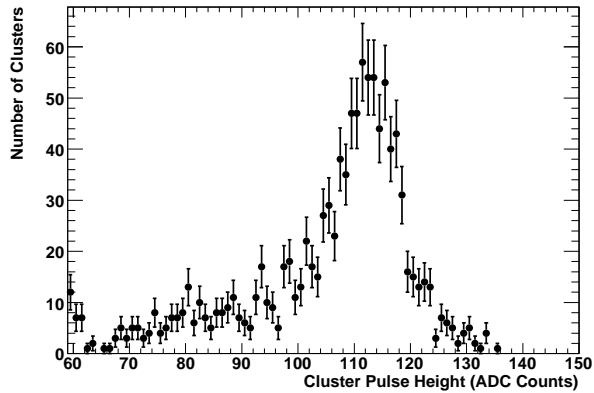, width=8.5cm,clip=}
\end{center}
\caption[]{$^{55}$Fe spectrum obtained on one analog output at room
temperature and for 6.25~MHz clock frequency.}
\label{fig:iron_spectrum}
\end{figure}

The sensor noise and uniformity among the different analog outputs are measured in the 
lab in dark. The pedestal level for each pixel is computed from the average pulse height
over 100 consecutive acquisitions. A pixel-to-pixel dispersion of less than 8~mV is obtained. 
The pedestal dispersion varies within 2\% among the different analog outputs of the same 
sensor. An average pixel leakage current below 10~fA is inferred from the results of 
measurements performed at different clock frequencies ranging from 25~MHz down to 6.25~MHz, 
corresponding to integration times of 2.6~ms and 10.5~ms, respectively. 
The sensor charge-to-voltage conversion gain is measured from the response
to 5.9~keV X-rays from a $^{55}$Fe source, corresponding to a charge generation of
1640 electrons. Figure~\ref{fig:iron_spectrum} shows the spectrum obtained on one 
analog output for a clock frequency of 6.25~MHz, at room temperature. A calibration 
of 14.9~e$^-$/ADC~count is obtained. The gain of the different analog outputs of the
same sensor is found to vary within 3\%. The charge-to-voltage conversion gain is expected 
to depend on the voltage applied to the pixel diode junction. For small signals, as those 
induced by one or few electrons, the capacitance of the junction can be considered constant and 
the conversion linear. For larger signals obtained when many electrons hit the same pixel in a 
single frame the capacitance of the diode junctions is expected to increase by several percent, 
up to $\sim$~20\% for signals corresponding to the full sensor dynamic range. Since most of the 
analyses reported in this paper deal with relatively small signals, up to few tens of electrons 
or less, we assume a constant calibration factor in the analysis. Typical average pixel noise 
figures of (32$\pm$2)~e$^-$ and (28$\pm$2)~e$^-$ of equivalent noise charge (ENC) are
measured under operating condition in the microscope set-up for clock frequencies of 25 and 
6.25~MHz, respectively.

\subsection{Electron Response}

\begin{figure}
\begin{center}
\begin{tabular}{c c}
\epsfig{file=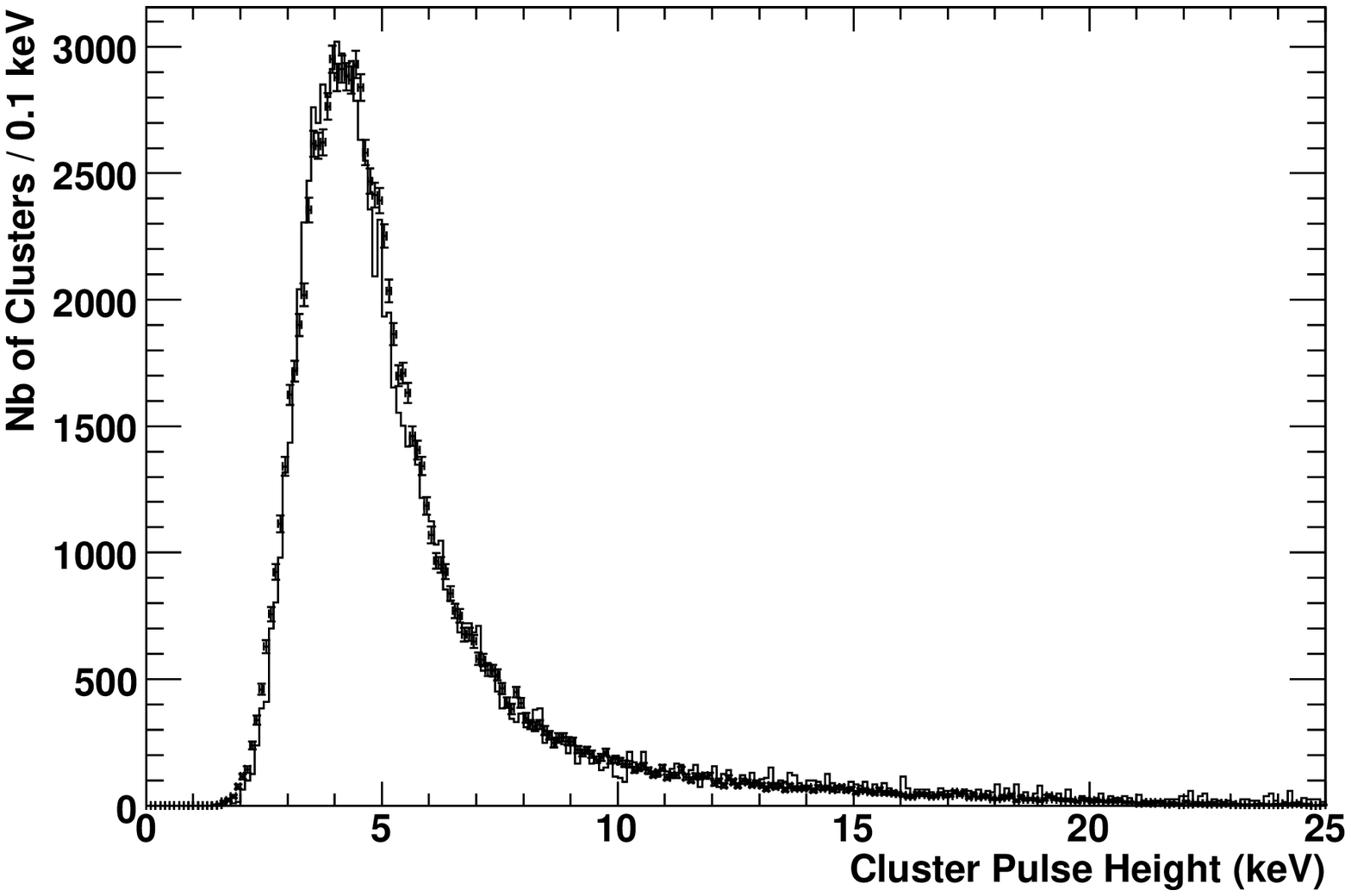,width=7.0cm,clip=} &
\epsfig{file=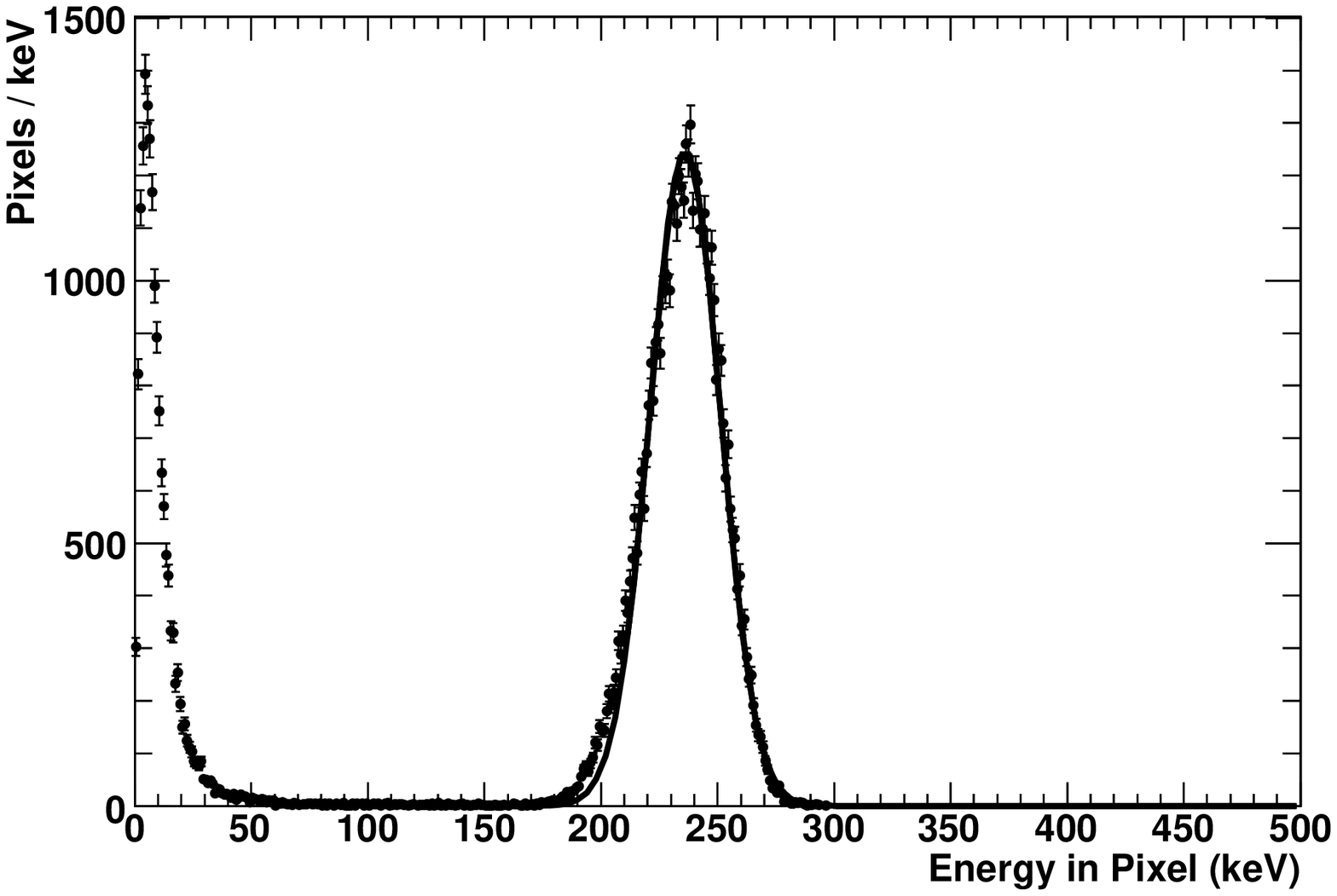,width=7.0cm,clip=} \\
\end{tabular}
\end{center}
\caption{Reconstructed deposited energy: (left) Energy recorded 
 in the 5$\times$5 pixel  matrix for single 300~keV electrons. 
 The points with error bars show the data and the histogram the 
 result of the {\tt Geant4}+{\tt PixelSim} simulation. (right) 
 Energy recorded per pixel for multiple electrons in bright 
 field illumination. The points show the data recorded in a single
 frame: pixels fully exposed to the electron beam exhibit a rather 
 uniform response described by a Gaussian distribution. Pixels with 
 low recorded energy correspond to the areas of the detector screened 
 with an Au wire and a metal plate knife edge (see text). Only pixels 
 with  at least 1~keV of recorded energy are shown.}
\label{fig:landau}
\end{figure}

The response to single electrons is characterised in terms of energy
deposition and cluster size.
We study the energy deposition in the sensor pixels, varying the number of 
electrons recorded per pixel and per frame. First, we operate with a flux of 
$\simeq$~50~$e^-$~mm$^{-2}$~frame$^{-1}$, which allows us to resolve 
individual electrons. Under these conditions, electron hits are reconstructed 
as pixel clusters. For these events the sensor response is characterised in 
terms of the number of pixels associated to a cluster and the pulse height 
measured in a 5$\times$5 pixel matrix centred around a seed pixel. The 
response measured on data is compared to the {\tt Geant4}+{\tt PixelSim} 
simulation. Figure~\ref{fig:landau} shows the reconstructed energy deposited
by 300~keV electrons for data and simulation. A good agreement is observed.

The uniformity across a detector sector has been studied for bright field 
illumination at a flux corresponding to an energy deposition of 
$\simeq$200~keV pixel$^{-1}$ frame$^{-1}$.
The distribution of the energy recorded in each pixel in a single frame is
shown in the right panel of Figure~\ref{fig:landau}. The dispersion of signal on 
the pixels fully exposed to the electron beam is obtained by a Gaussian fit to the 
recorded energy distribution. The relative r.m.s.\ dispersion at 300, 200, 120 and 
80~keV is 0.063, 0.058, 0.080 and 0.11, respectively. The increase at lower energies 
is interpreted as an effect of the larger energy loss fluctuations, since measurements 
are performed at constant energy per pixel instead of constant number of electrons per 
pixel. 

\begin{figure}
\begin{center}
\begin{tabular}{c c}
\epsfig{file=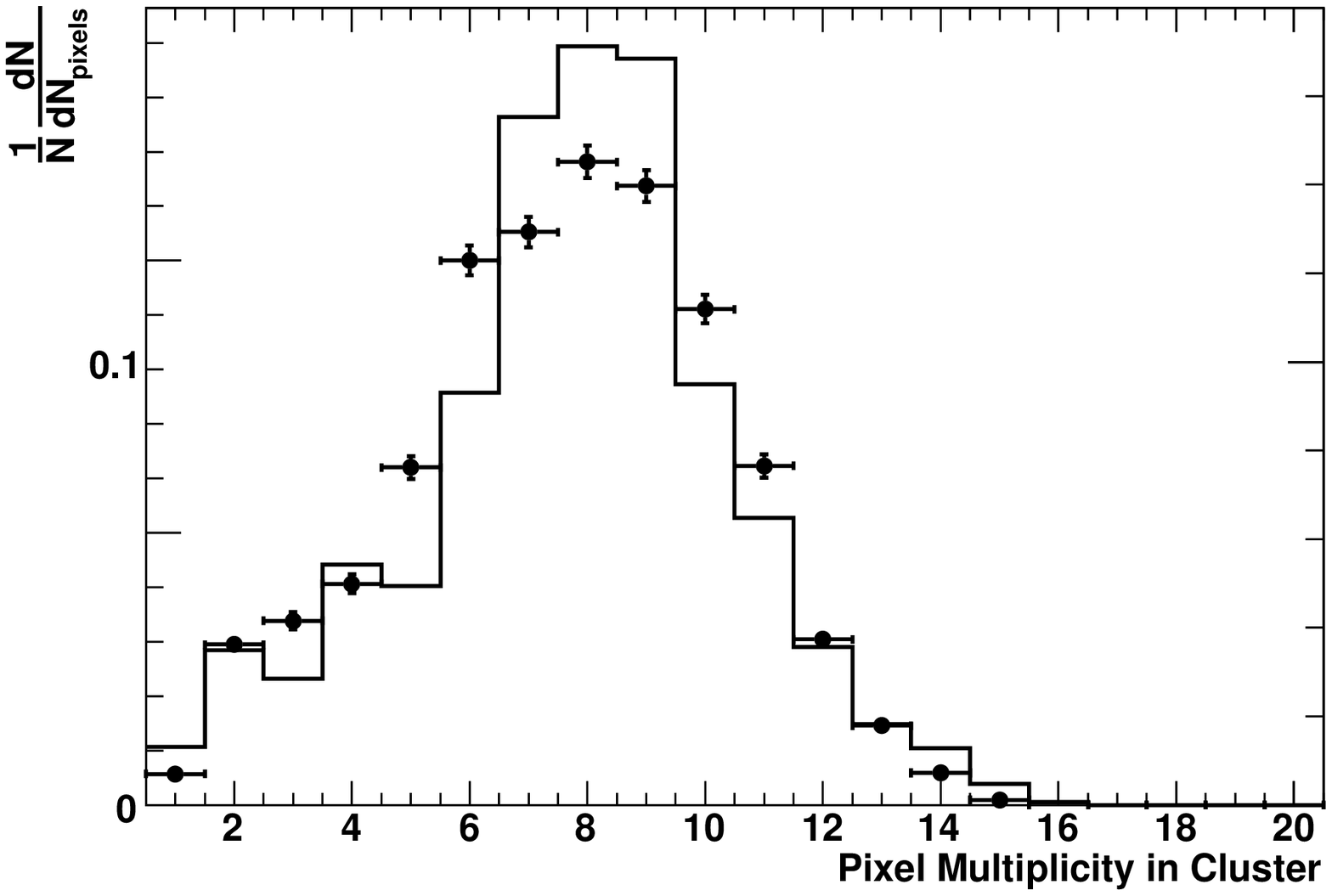,width=6.0cm,clip=} &
\epsfig{file=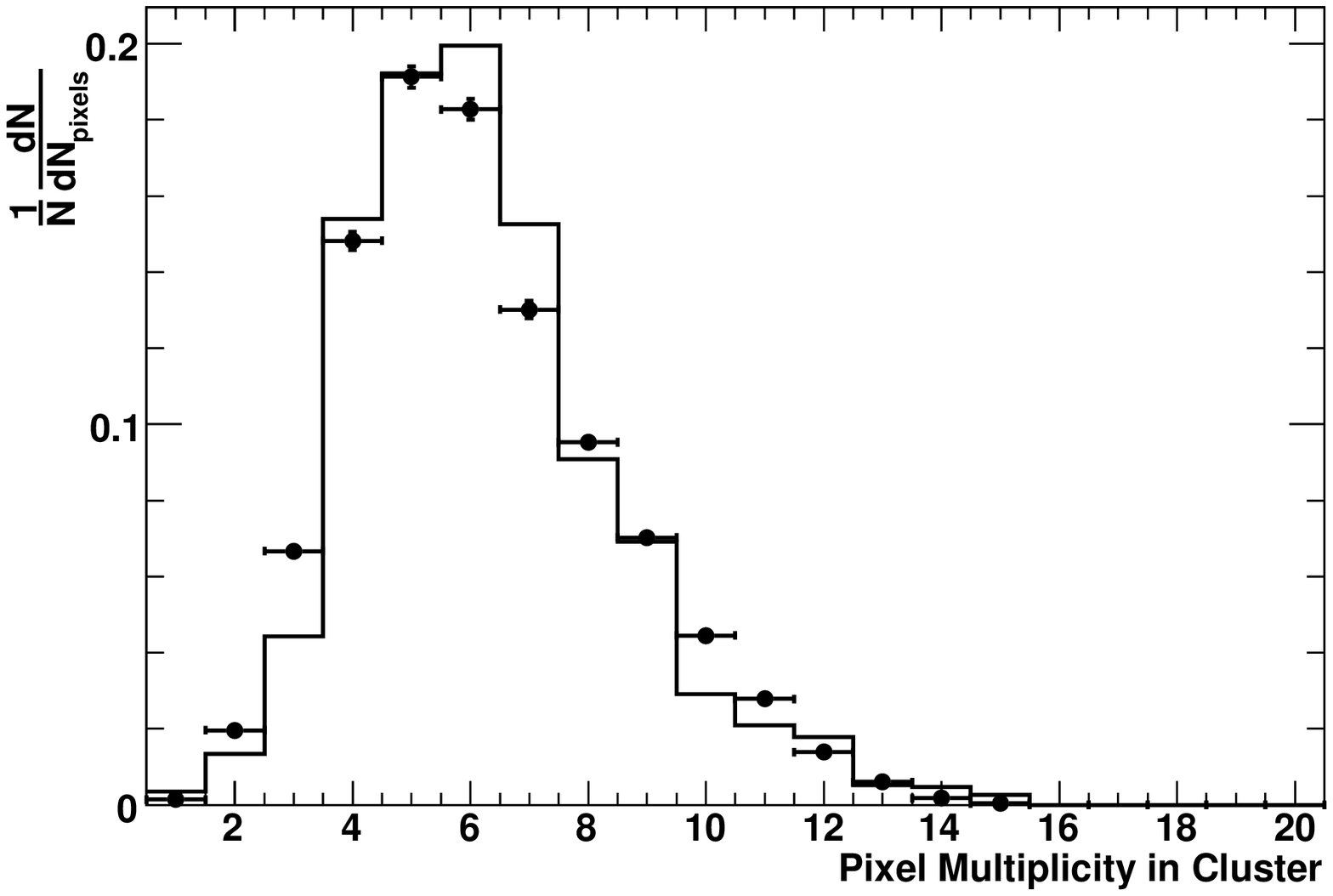,width=6.0cm,clip=} \\
\epsfig{file=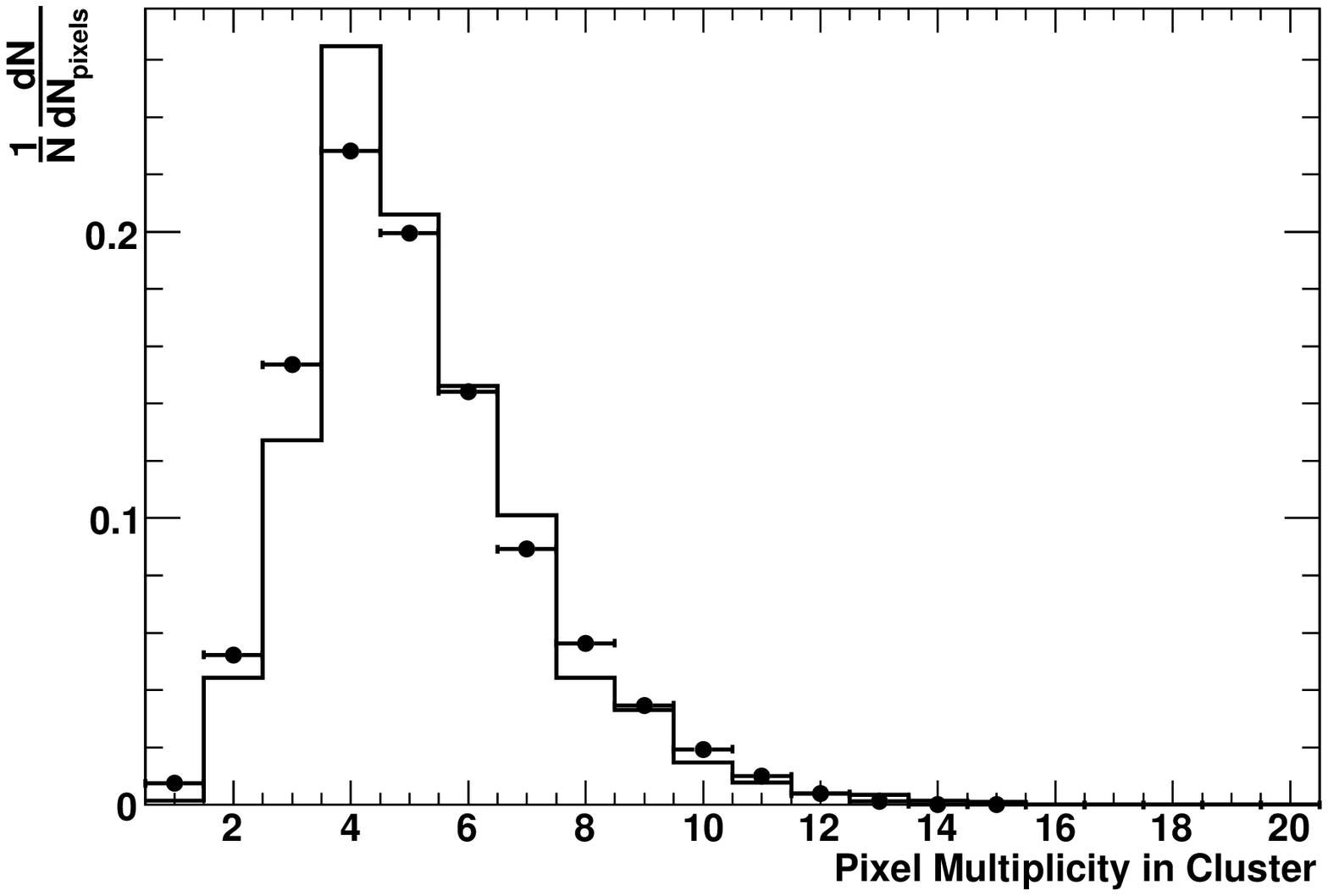,width=6.0cm,clip=} &
\epsfig{file=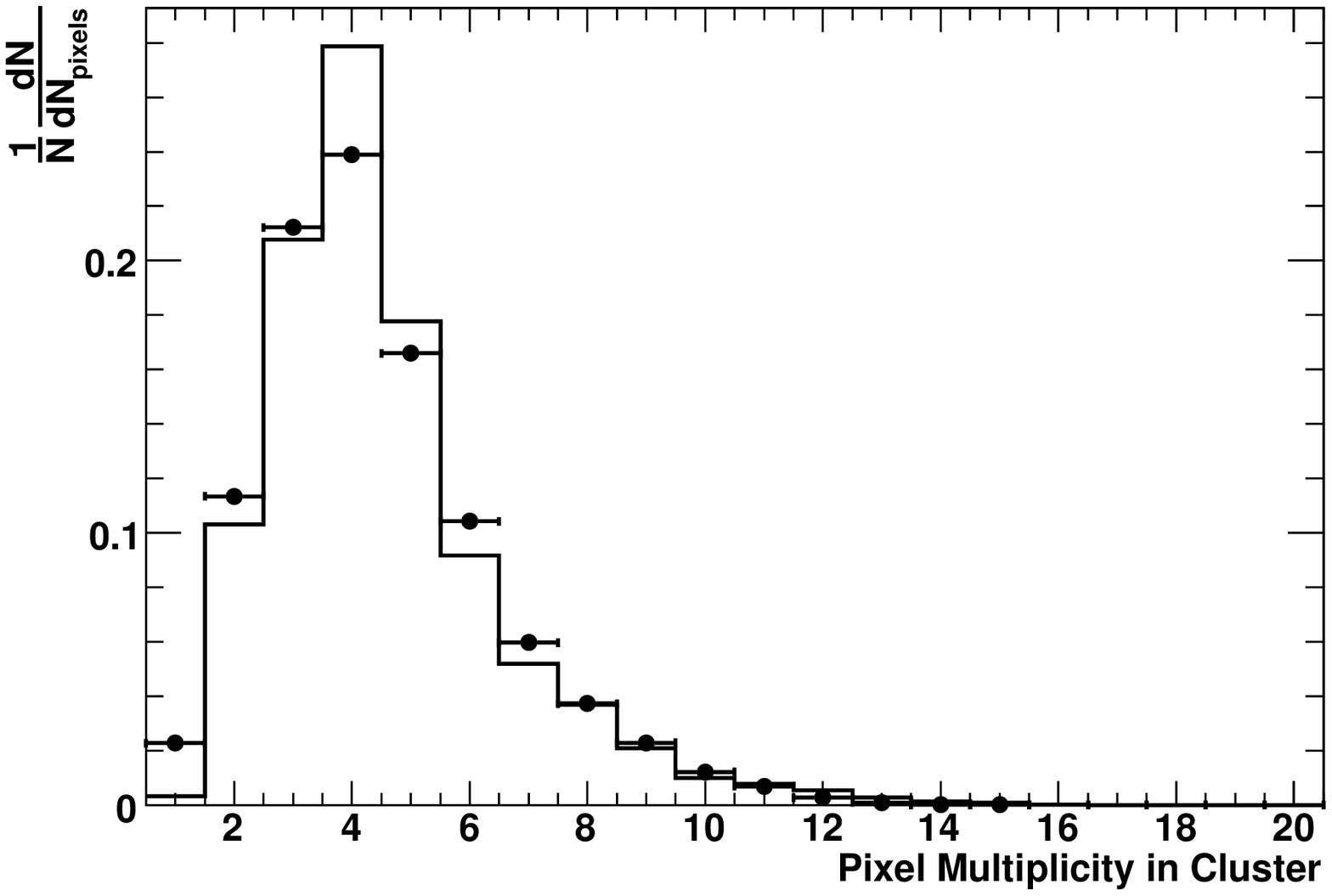,width=6.0cm,clip=} \\
\end{tabular}
\end{center}
\caption{Number of pixels in single electron clusters at various 
electron energies: 80~keV (upper left panel),  150~keV (upper right 
panel), 200~keV (lower left panel) and 300~keV (lower right panel).
Points with error bars represent the data and the line the result of 
the {\tt Geant4}+{\tt PixelSim} simulation.}
\label{fig:npixels}
\end{figure}

\begin{figure}
\begin{center}
\epsfig{file=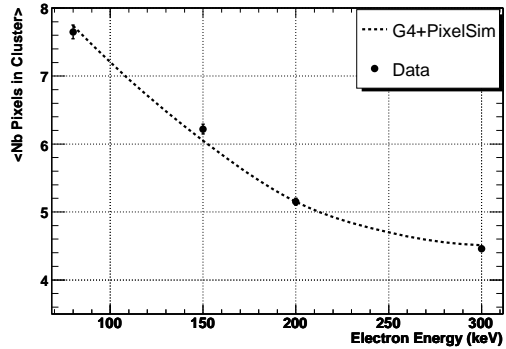,width=7.5cm,clip=} 
\end{center}
\caption{Average pixel multiplicity vs.\  electron energy.  The points with 
error bars show the data and the line the result of the 
{\tt Geant4}+{\tt PixelSim} simulation.}
\label{fig:npixelsE}
\end{figure}

The study of the cluster size allows us to investigate the charge spread 
around the point of impact of each electron onto the detector. This is used 
to validate the simulation and to assess the variation of the charge spread 
as a function of the electron energy. A clustering algorithm with two thresholds 
is used to determine the cluster size~\cite{Battaglia:2008yt}. 
First the detector is scanned for ``seed'' pixels with pulse height values 
over a S/N threshold set to 3.5. Seeds are sorted according to their pulse heights 
and the surrounding neighbouring pixels are added to the cluster if their S/N 
exceeds 2.5. In simulation, the $\sigma_{\mathrm{diff}}$ parameter is tuned to 
minimise the $\chi^2$ of the pixel multiplicity distribution in the clusters for 
simulation vs.\ data at 300~keV, where multiple scattering is lower, 
as discussed above.  The pixel multiplicities in electron clusters at various energies 
are shown in Figure~\ref{fig:npixels} and the evolution of the average pixel multiplicity 
vs.\ electron energy in Figure~\ref{fig:npixelsE}. The agreement of the tuned {\tt PixelSim} 
simulation with data is good and we observe an increase of the pixel multiplicity due to 
the increased multiple scattering and energy deposition at lower energies.

\section{Imaging Characterisation}

The pixel imaging performance is characterised in terms of three observables: the line spread 
function (LSF), the modulation transfer function (MTF) and the detection quantum 
efficiency (DQE). 
We study these observables for two different imaging regimes, bright field 
illumination, where each pixel typically receives several electrons per 
frame, and for cluster imaging, a recently proposed alternative imaging 
technique~\cite{Battaglia:2009dt}, where the electron flux is kept low enough that 
the position of impact of each individual electron is reconstructed by interpolating 
the charge collected on the pixels of a signal cluster. This allows us to achieve 
spatial sampling with frequencies much larger than the Nyqvist frequency.

The point spread function is one of the key features for an imaging sensor in 
transmission electron microscopy. It depends on several parameters of which the 
most important are pixel size, electron multiple scattering in the active layer and 
in its vicinity and charge carrier diffusion. For these tests the detector is mounted 
on a proximity board which is cut below its active area to eliminate back-scattering 
from the board material. Pixel sensors of both 300~$\mu$m- and 50~$\mu$m-thickness are 
tested. A 60~$\mu$m-diameter Au wire is mounted parallel to the pixel rows on top of 
each sensor at a distance of $\sim$2~mm above its surface.

Another important feature is the image contrast, defined by the ratio between 
the signal on the pixels directly exposed to the beam to the response of those which 
are covered.  The contrast depends on charge leakage, scattering and noise. We measure it 
using both the wire and a metal foil covering the upper portion of the chip. We expect 
the contrast to improve for thin sensors, where the contribution of electrons back-scattered 
in the the bulk Si and depositing energy in the shadowed area is smaller.

\subsection{Bright Field Illumination}

\begin{figure}
\begin{center}
\begin{tabular}{c c}
\epsfig{file=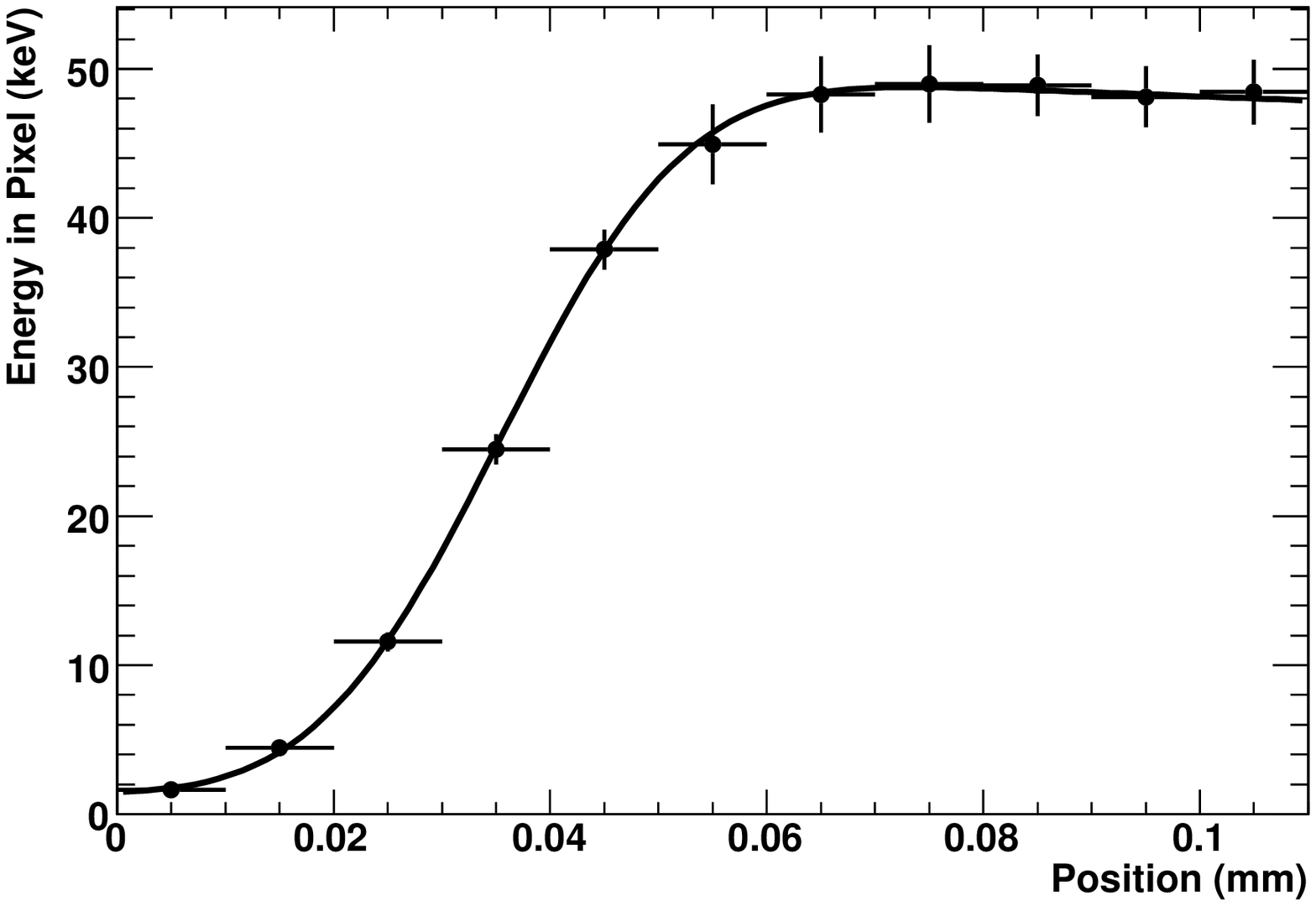,width=7.0cm,clip=} &
\epsfig{file=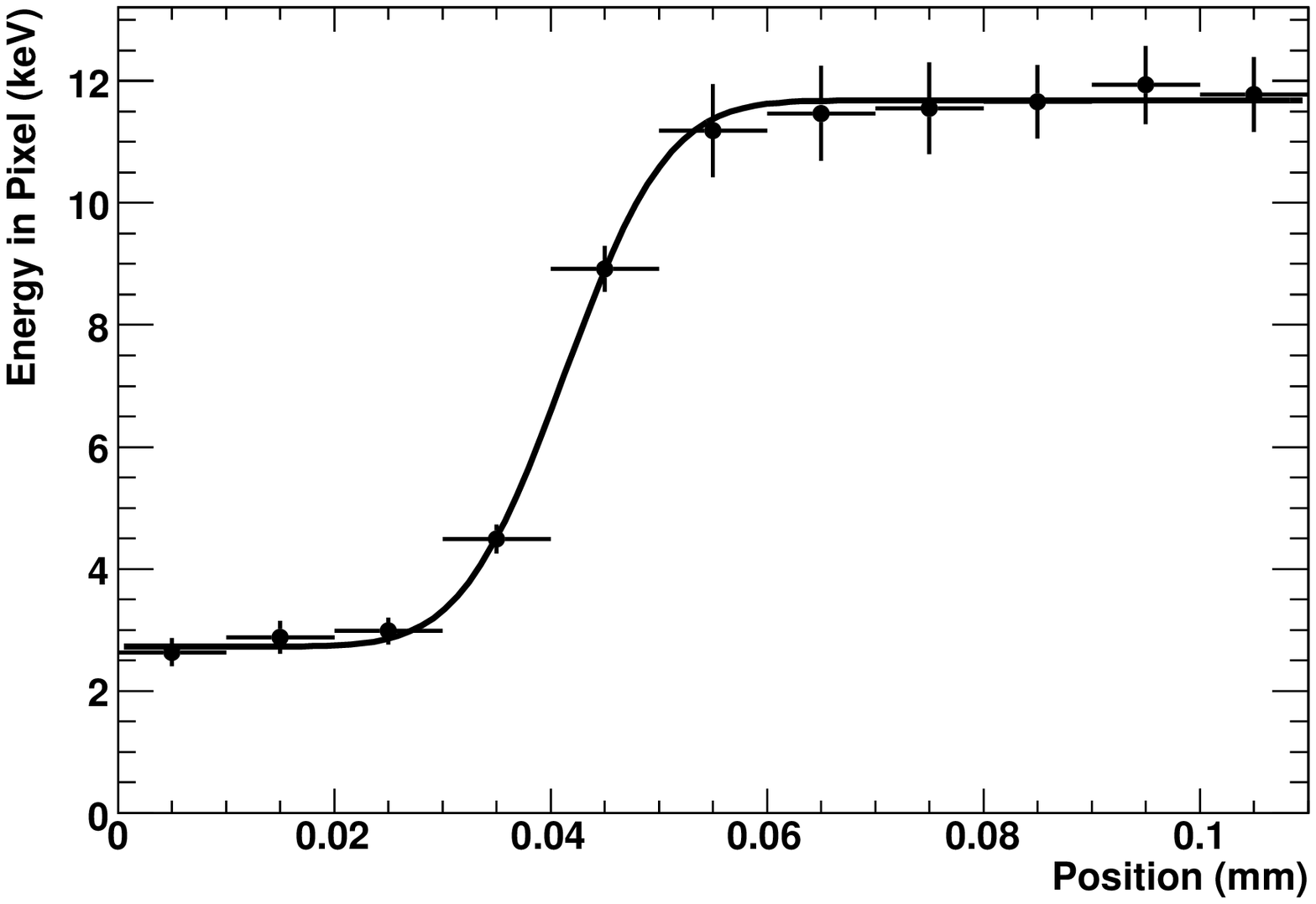,width=7.0cm,clip=} \\
\end{tabular}
\end{center}
\caption{Pulse heights measured on pixels along a row across the Au wire stretched 
 above the pixels for 80~keV (left panel) and 300~keV (right panel) beam. 
 Data are shown as points with error bars while the continuous line shows 
 the sigmoid function corresponding to the best fit.}
\label{fig:LSF}
\end{figure}

In simulation, a monochromatic, point-like beam of electrons is sent onto the 
surface of the detector. The line spread function (LSF) is determined as the r.m.s.\ 
of the predicted distribution of the detected charge on the pixels.
On the data the LSF for bright field illumination is determined using the image 
projected onto the sensor by the thin Au wire, using the same technique as 
in~\cite{Battaglia:2008yt}. Since the gold wire has well-defined edges,
the profile of the deposited energy in the pixels, measured across the wire, allows 
us to study the charge spread due to scattering and diffusion along the projected 
image of the wire edge and compare to simulation.  We study the change in the recorded 
signal, by scanning along pixel rows across the gold wire. We extract the LSF from 
images such as those in Figure~\ref{fig:LSF}, which show the pulse heights measured 
on the pixels along a set of rows. 
In~\cite{Battaglia:2008yt}, we parametrised the measured pulse height on pixel rows across 
the wire image with a box function smeared by a LSF Gaussian term and extracted the LSF 
by a 1-parameter $\chi^2$ fit with the Gaussian width as free parameter. Here we adopt an 
extension of this method which uses the sum of several sigmoid functions to 
describe the edge as proposed in~\cite{optik}:
\begin{eqnarray}
PH = a_0 + \sum_i a_i Erf\big(\frac{x_0-x}{\sqrt{2} \sigma_i} \big)
\label{eq:lsf}
\end{eqnarray}
where the error function is defined as $Erf(x) = \frac{2}{\sqrt{\pi}} \int_{0}^{x} dt~e^{-t^2}$.
We find that one or two sigmoid functions are sufficient to describe the data.
We use a single sigmoid function to obtain LSF values of (7.6$\pm$0.6)~$\mu$m at 300~keV 
and (12.6$\pm$0.7)~$\mu$m at 80~keV, where the quoted uncertainty is statistical. These results 
are consistent with those obtained from simulation and with the data in our previous study.
Since it is customary to quote the imaging resolution in terms of the modulation transfer 
function (MTF), which is the Fourier transform of the line spread function, we repeat the 
fit using the sum of two sigmoid functions in Eq~\ref{eq:lsf}.
\begin{figure}
\begin{center}
\epsfig{file=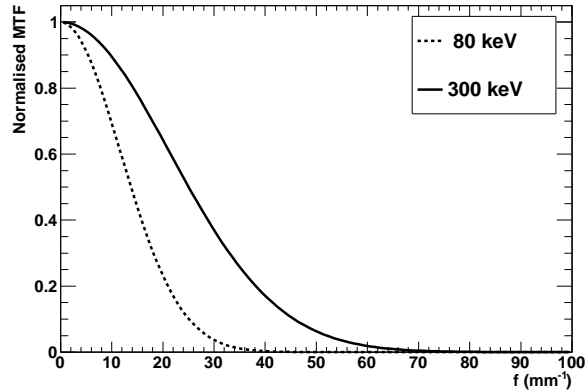,width=8.5cm,clip=}
\end{center}
\caption{Normalised modulation transfer function obtained on data for bright field 
illumination at 80~keV (dotted line) and 300~keV (continuous line).}
\label{fig:MTF}
\end{figure}
The width of the first, 
$\sigma_1$, is fixed to its value obtained in the previous fit, using a single function, while 
the width of the second function, $\sigma_2$, as well as the two normalisation coefficients $a_1$ 
and $a_2$ are left free. We observe that the $\chi^2$ at the minimum improves only marginally 
by performing the fit with two functions compared to a single function, as used to extract the 
LSF. Figure~\ref{fig:MTF} shows the MTF obtained with this method on data, for 80~keV and 300~keV 
electrons.

\begin{figure}
\begin{center}
\begin{tabular}{c c}
\epsfig{file=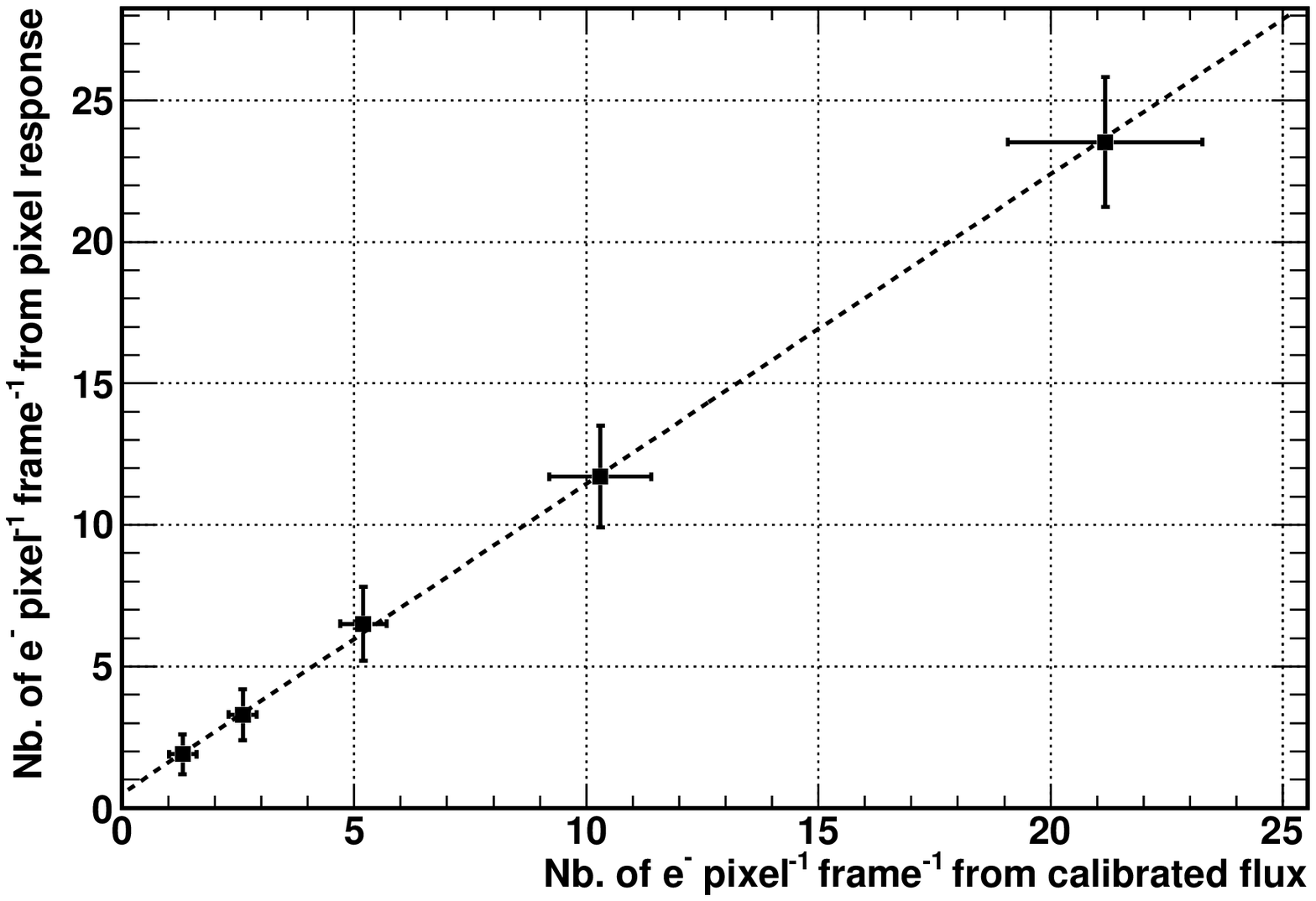,width=6.5cm,clip=} &
\epsfig{file=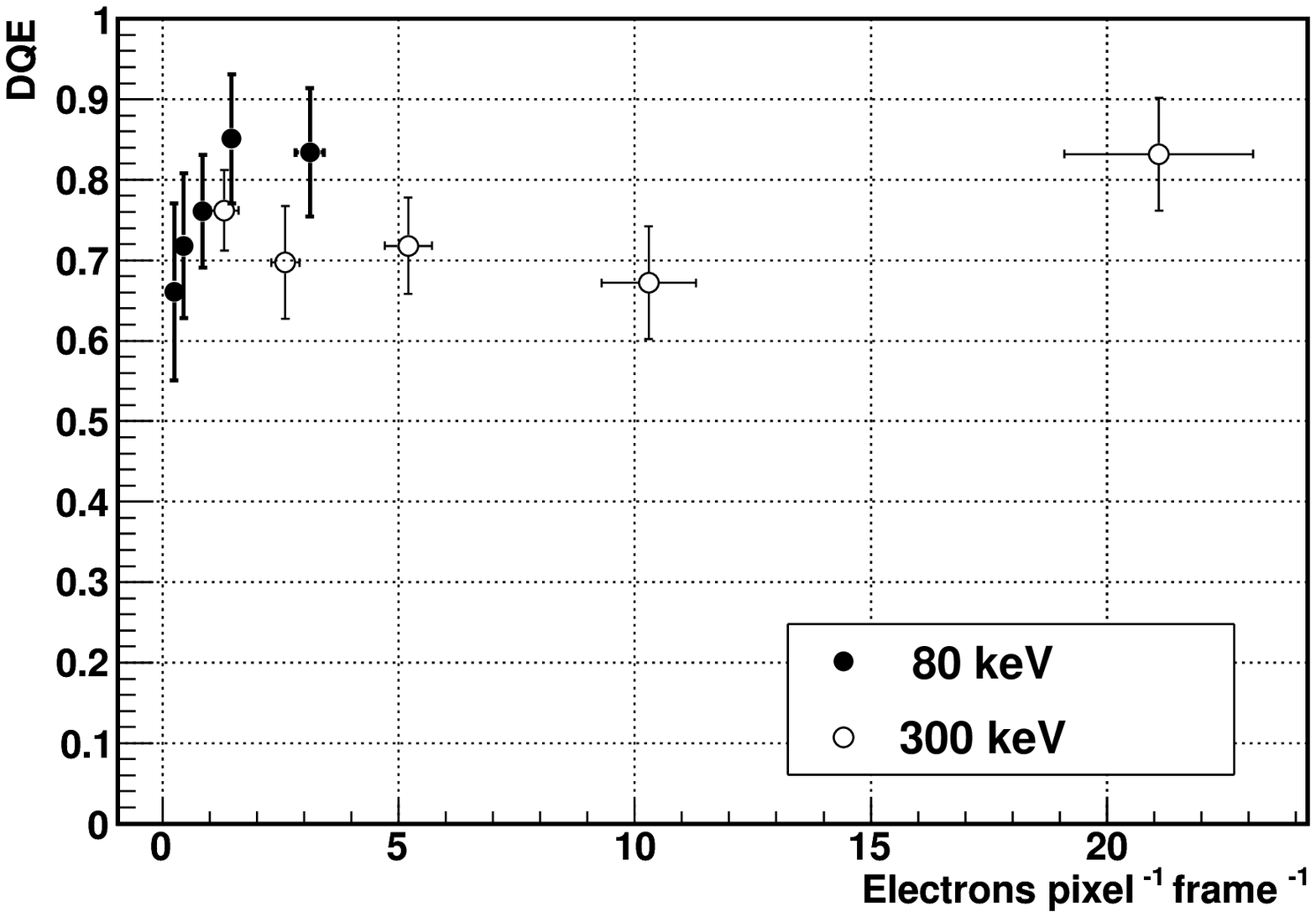,width=7.0cm,clip=} \\
\end{tabular}
\end{center}
\caption{Reconstructed electron flux from pixel response and DQE: (left) Number of electrons 
per pixel and frame obtained from the measured energy in the pixels after unfolding the contribution 
of charge spread as a function of the result from the electron flux obtained from the beam 
current read on the microscope calibrated screen at 300~keV. The fitted line has a slope of 
1.09$\pm$0.11. (right) Measured DQE as a function of the electron flux at 80~keV and 300~keV.}
\label{fig:plotNe}
\end{figure}

Next we determine the detection quantum efficiency, a quantity widely used 
to characterise electron detection systems, introduced in~\cite{dqe}. We compute 
the DQE accounting for the charge migration between neighbouring pixels due to 
charge carrier diffusion following the method adopted in~\cite{ishizuka,horacek}: 
\begin{eqnarray}
DQE \equiv \frac{(S/N)^2_{out}}{(S/N)^2_{in}} = \frac{(S/N)^2_{out}}{m \times N}
\label{eq:dqe}
\end{eqnarray}
where $N$ is the number of electrons pixel$^{-1}$ frame$^{-1}$ and $m$ is the mixing 
factor, given by $m = 1/\sum f_{ij}^2$, with $f_{ij}$ being the fraction of the charge 
collected at pixel $j$ for an electron hitting the detector on pixel $i$. We extract the 
$f_{ij}$ coefficients by studying the charge distribution in the signal clusters for 
single electrons, as discussed in section 4.3.
We perform this measurement varying the electron flux and measuring the current on the 
calibrated phosphor screen of the Titan microscope. We take data at different fluxes 
in the range 1.3 to 21~e$^-$ pixel$^{-1}$ frame$^{-1}$ at 300~keV and 
0.3 to 3.1~e$^-$ pixel$^{-1}$ frame$^{-1}$ at 80~keV. First, we determine the number of 
electrons per pixel and per frame from the pixel single electron response in terms of 
pulse height and charge spread, discussed in section~4.3. 
The left panel of Figure~\ref{fig:plotNe} shows the number of electrons per pixel and 
frame measured from the pixel response as a function of that from the beam current
on the phosphor screen, which exhibits a linearity within $\simeq$~10~\%.
We measure the DQE at different fluxes using Eq.~(\ref{eq:dqe}). 
Results are consistent within the statistical uncertainties (see the right panel of 
Figure~\ref{fig:plotNe}). By averaging these results, we obtain DQE = 0.74$\pm$0.03 
at 300~keV and 0.78$\pm$0.04 at 80~keV, where the quoted uncertainties are statistical.

\subsection{Cluster Imaging}

In bright field illumination the point spread function has a contribution from the
lateral charge spread due to charge carrier diffusion in the active detector volume. 
At high rate, the signal recorded on each individual pixel is the superposition of 
the charge directly deposited by a particle below the pixel area with that collected 
from nearby pixels through diffusion, multiple scattering and back-scattering from the 
bulk Si. If the electron rate is kept low enough so that individual electron clusters 
can be reconstructed, the position of passage of each electron can be obtained from 
the centre of gravity of the observed signal charge. The leakage of charge on neighbouring 
pixels is taken into account through the centre of gravity calculation and the precision
depends on the signal-to-noise ratio.  This technique, widely adopted in 
tracking applications for accelerator particle physics, provides us with a significant 
gain in spatial resolution. In a recent letter to this journal~\cite{Battaglia:2009dt} 
we proposed to adopt the same technique for imaging, with low electron fluxes, 
i.e. $\le$~10$^2$~e$^-$~mm$^{-2}$~frame$^{-1}$. We named this technique ``cluster imaging''. 
In this regime, the image is formed by adding many subsequent frames. We showed that for 
cluster imaging the LSF depends only on the detector pixel size and cluster S/N 
(determining the single point resolution) and on the multiple scattering. A significant 
improvement in the LSF can be obtained.
\begin{figure}
\begin{center}
\begin{tabular}{c c}
\epsfig{file=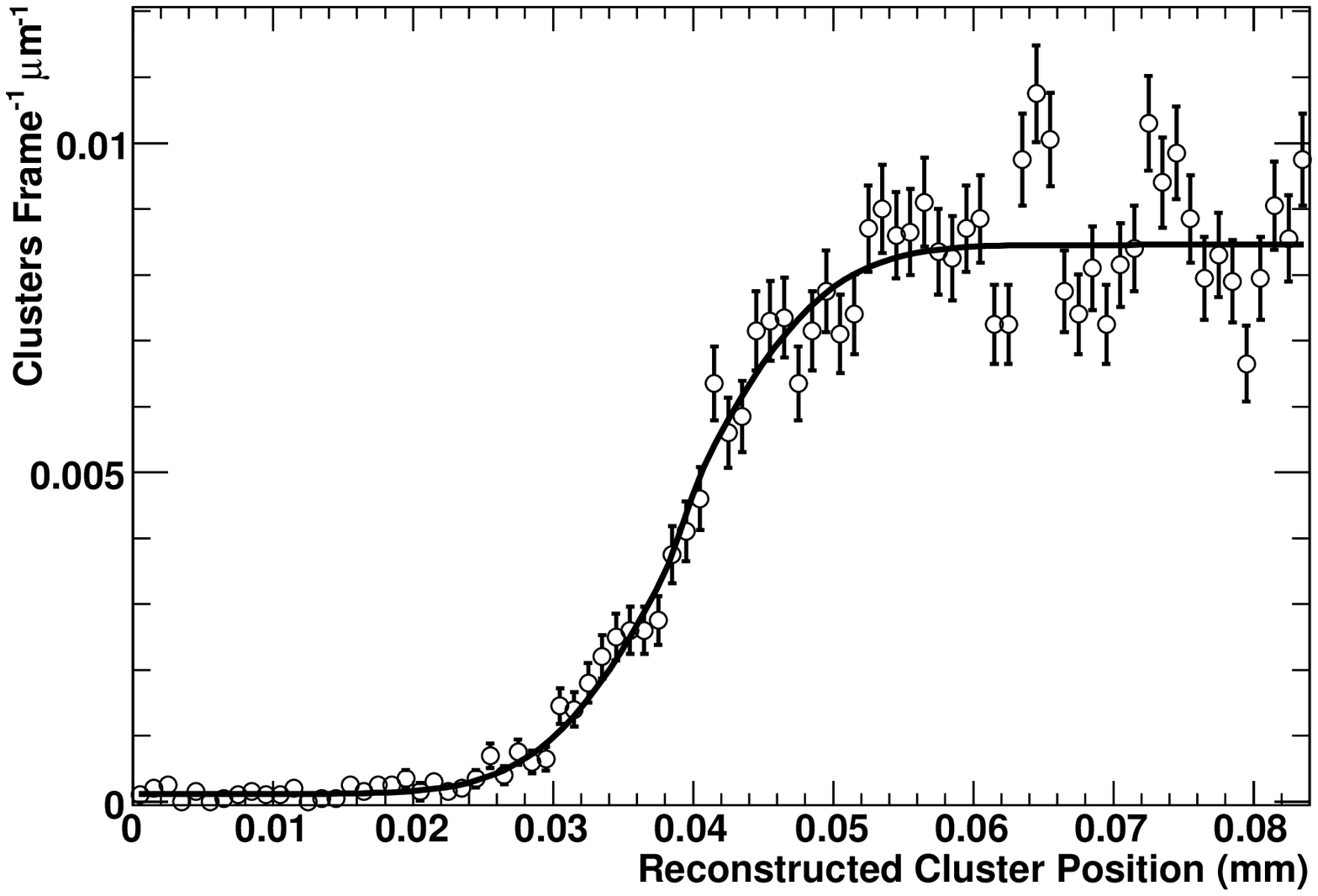,width=7.0cm,clip=} &
\epsfig{file=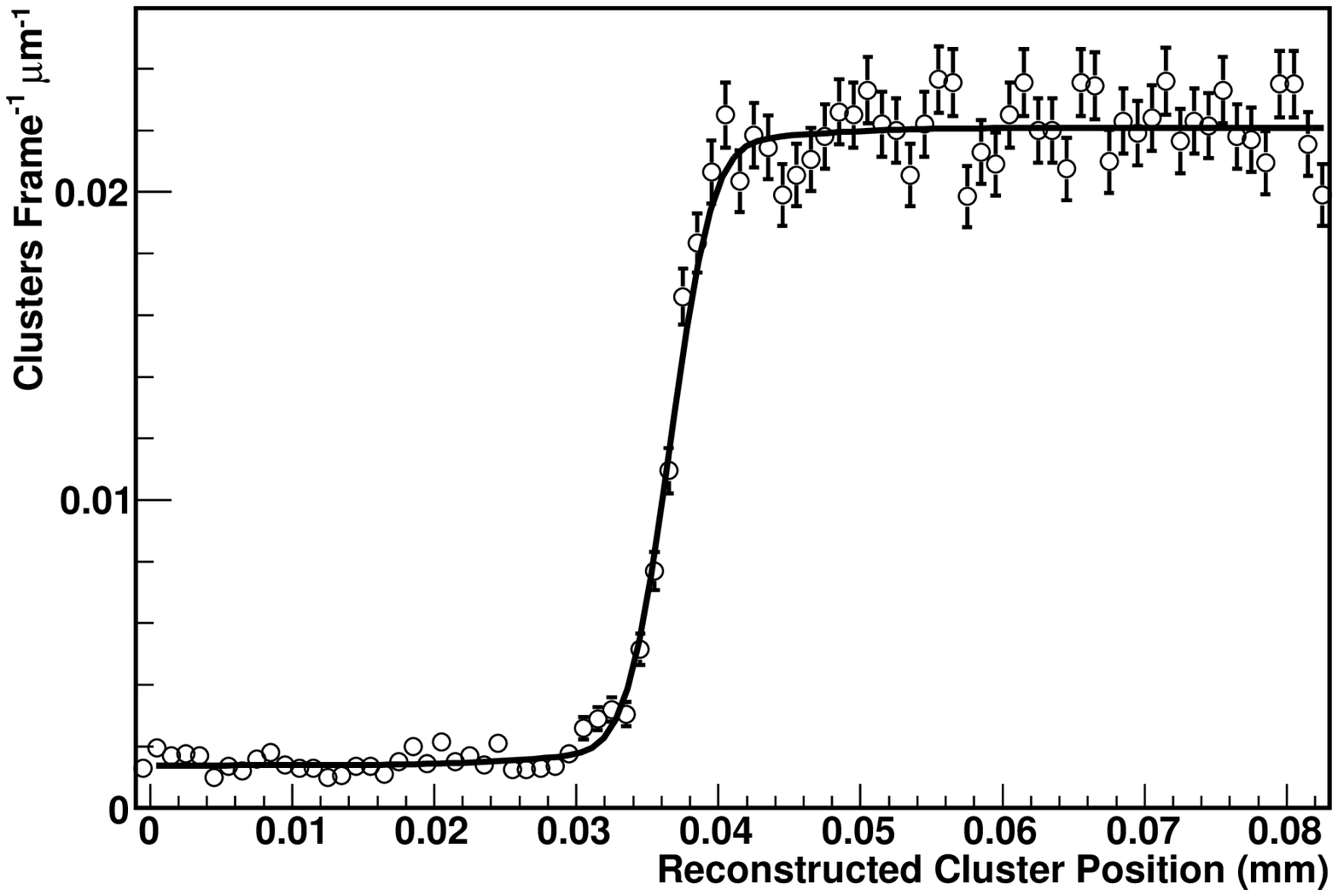,width=7.0cm,clip=} \\
\end{tabular}
\end{center}
\caption{Pulse heights measured on pixels along a row across the Au wire stretched 
 above the pixels for 80~keV (left panel) 300~keV (right panel) beam. 
 Data are shown as points with error bars while the continuous line shows 
 the sigmoid function corresponding to the best fit.}
\label{fig:LSFci}
\end{figure}
\begin{table}
\caption{Line spread function values measured with bright field illumination and cluster 
imaging at various electron energies.}
\begin{center}
\begin{tabular}{|l|c|c|}
\hline
Electron    & Bright Field    &  Cluster Imaging \\
Energy      & LSF             &  LSF            \\
(keV)       & ($\mu$m)        &  ($\mu$m)        \\
\hline 
~80         &  12.1$\pm$0.7  &  6.68$\pm$0.34    \\
150         &  11.2$\pm$0.6  &  4.64$\pm$0.27    \\
200         &   9.4$\pm$0.6  &  3.53$\pm$0.16    \\
300         &  ~7.4$\pm$0.6  &  2.35$\pm$0.15    \\
\hline
\end{tabular}
\end{center}
\label{tab:LSF}
\end{table}
\begin{figure}
\begin{center}
\epsfig{file=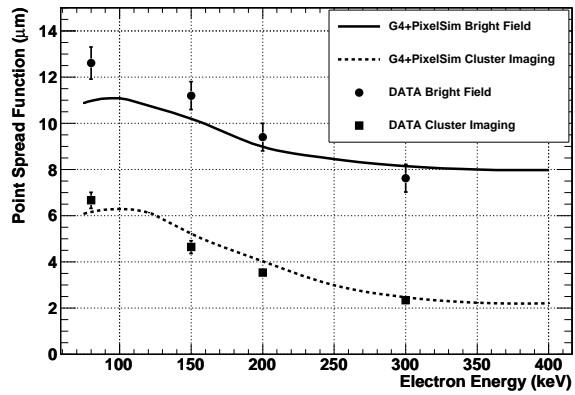,width=8.5cm,clip=}
\end{center}
\caption{Fitted line spread function as a function of beam energy. 
 Cluster imaging (squares with error bars) and  bright field illumination
 (circles with error bars) data are compared to {\tt Geant4}+{\tt PixelSim} 
 simulation.} 
\label{fig:plotPSFCI}
\end{figure}
Here, we repeat the analysis by extracting the LSF from the wire edge images built by the 
superposition of multiple frames taken with $\simeq$~10-40~e$^-$~mm$^{-2}$~frame$^{-1}$,
using the same sigmoid fit adopted above. 
\begin{figure}
\begin{center}
\epsfig{file=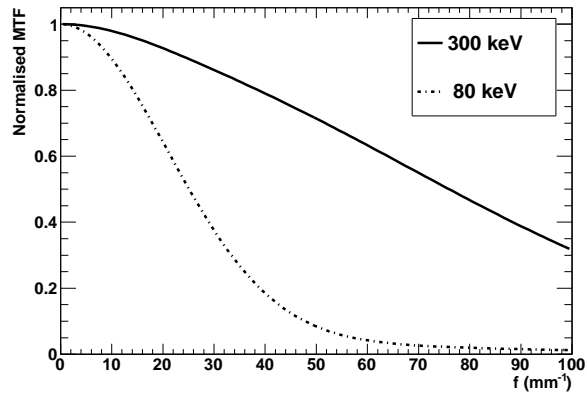,width=8.5cm,clip=}
\end{center}
\caption{Normalised modulation transfer function obtained on data for cluster imaging 
illumination at 80~keV (dotted line) and 300~keV (continuous line).}
\label{fig:MTFci}
\end{figure}
First we use a single sigmoid to extract the 
value of the LSF. Results are summarised in Table~\ref{tab:LSF} and Figure~\ref{fig:plotPSFCI}, 
where the values for bright field illumination and cluster imaging measured at different 
energies are compared.
Then, we fix the width of the first function to this fitted value and we introduce a 
second sigmoid with free width and fit it together with the normalisation coefficients, 
as discussed above. The LSF function, parametrised according to Eq.~(\ref{eq:lsf}), 
is used to extract the MTF. Results are shown in Figure~\ref{fig:MTFci} for 300~keV 
and 80~keV electrons.

\subsection{Image Contrast and Imaging Tests}

We study the contrast ratio obtained for a 50~$\mu$m-thick sensor and a sensor
which has the original 300~$\mu$m Si thickness, by comparing the signal observed 
on the pixel area covered by the wire to that obtained on the pixels directly exposed 
to the beam. We measure a ratio of 2.95 for the 300~$\mu$m-thick 
and 4.49 for the 50~$\mu$m-thick sensor, using bright field illumination.
\begin{figure}[hb!]
\begin{center}
\epsfig{file=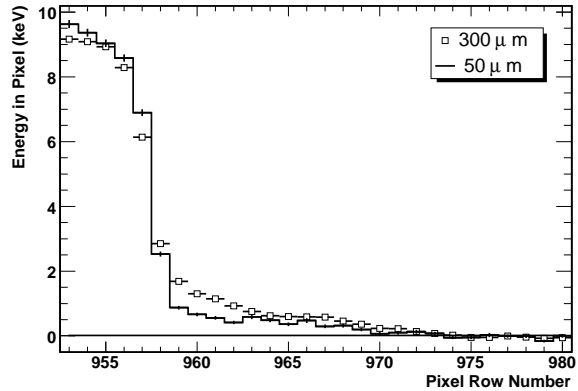,width=8.5cm,clip=}
\end{center}
\caption{Pulse heights measured on pixels along rows across a metal knife edge for 300~keV beam
with a 300~$\mu$m (squares with error bars) and a 50~$\mu$m (line with error bars) thick 
TEAM1k sensor. The enhanced leakage of charge in the pixels below the metal plate due to electron 
scattering in the thick sensor is evident.}
\label{fig:edge}
\end{figure} 
The contrast is affected by multiple scattering effects in the detector and the 
thinned sensor performs better due to the reduced back-scattering in the bulk 
Si underneath the sensitive epitaxial layer. The contrast decreases at lower 
energies, we measure 3.9 on the thin sensor at 150~keV, until the range of the 
electrons becomes short enough that the scattering contribution to the energy 
leaking under the shadow of the wire becomes small and a contrast ratio in 
excess of 20 is measured at 80~keV. Then, we study the pixel response at the edge 
of a metal plate on the thick and the thinned sensor. The plate covers an area where 
pixels do not receive direct electron hits. Figure~\ref{fig:edge} shows the pulse 
height measured on pixels along rows across the plate edge. Beyond the sharp edge 
we observe that the recorded signal in the thick sensor falls less 
rapidly compared to that in the thin sensor, which we interpret as an effect of 
charge deposited by back-scattered electrons. We also measure the LSF by fitting a 
single sigmoid function to the pixel response. We obtain (10.4$\pm$0.5)~$\mu$m and 
(7.7$\pm$0.4)~$\mu$m for the 300~$\mu$m- and the  50~$\mu$m-thick sensor, respectively.
The latter result is compatible to that obtained with the thin Au wire.
\begin{figure}
\begin{center}
\epsfig{file=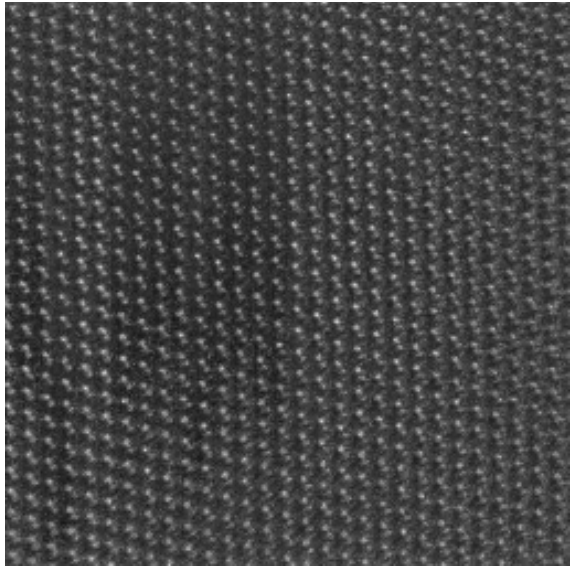,width=7.5cm,clip=}
\end{center}
\caption{Raw image of a single 2.6~ms exposure of a Si sample in $<110>$ orientation taken with the 
TEAM1k chip and a 300~keV beam at the TEAM1 microscope. The image has no corrections applied, which 
gives an impression of the overall uniformity across the detector.}
\label{fig:si}
\end{figure}
Finally, we perform an imaging test using a Si sample on the TEAM1 microscope with 300~keV electrons. 
Figure~\ref{fig:si} gives a qualitative demonstration of the performance of the TEAM1k detector. 
The image shows a single 2.6~ms exposure revealing the Si dumbbells close to the $<110>$ zone axis, 
with a spacing of 1.36~Angstrom.

\section{Sensor irradiation}

The radiation tolerance of the chip has been assessed by irradiating a TEAM1k sensor
with 300~keV electrons, up to a dose of 5~Mrad. The damage mechanism is an increase of
leakage current, which not only increases noise, but also decreases the dynamic range. CMOS 
active pixel sensors are capable of good S/N because the charge collection node capacitance
is small, resulting in a high conversion gain (V/$e^-$) and thus high sensitivity to leakage
current.  The irradiation has been performed 
in subsequent steps with a flux of 875~$e^-$s$^{-1}\mu$m$^{-2}$, corresponding
to a dose rate of 250~rad~s$^{-1}$, from the measured beam current on a known beam spot area onto 
the Titan calibrated phosphor screen. In-between consecutive
irradiation steps, several dark frames are acquired without beam, in order to follow 
the evolution of the pixel leakage current with the dose, and with very low intensity
beam, to monitor the pixel response to single electrons and the gain calibration.
A small portion of the sensor active surface is covered with a gold wire, as 
described in Section~5. Bright field images of this wire are acquired throughout 
the irradiation in order to monitor also the sensor imaging performance.
All tests are performed with the detector cooled at +5$^{\circ}$C. 
The sensor is glued with thermally conductive epoxy to an AlN
substrate board, and wire-bonded to a flex circuit glued on top of the AlN board.
Heat is removed from the sensor by a copper finger that contacts the AlN board
and is cooled by a double-Peltier system.

\begin{figure}[h]
\begin{center}
\epsfig{file=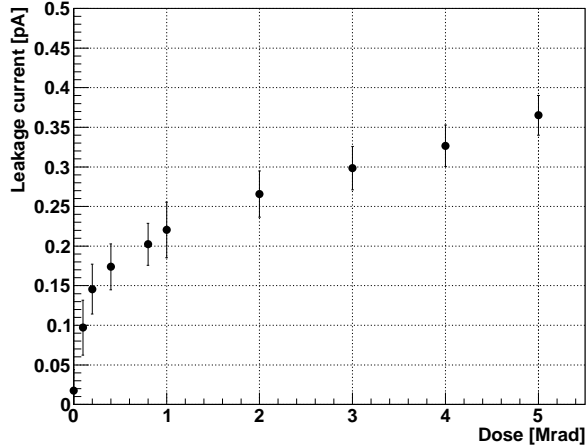, width=8.5cm}
\end{center}
\caption[]{Leakage current as a function of the 300~keV electron dose.}
\label{fig:leakage}
\end{figure}

Figure~\ref{fig:leakage} shows the evolution of the pixel leakage current with dose.
The leakage current is determined by comparing the pixel base level at two different
clock frequencies, 25~MHz and 6.25~MHz, corresponding to integration times of 
2.6~ms and 10.5~ms, respectively. A sub-linear increase of the current with
dose is observed, reaching about 0.4~pA after 5~Mrad. The
leakage current increase is due to ionising damage in the field oxide that leads to
trapping of positive charge. The inversion of the Si interface results in an increased 
current in the charge collecting diode. 300 keV electrons are not expected to 
significantly affect the silicon bulk through displacement damage.

Results of the irradiation test of an earlier prototype chip with a comparable pixel 
cell manufactured in the same process gave a leakage current of 0.2~pA after a dose of 
1.11~Mrad with 200~keV electrons on 20$\times$20~$\mu$m$^2$ pixels. The test was performed 
at room temperature and for an integration time of 737~$\mu$s. Considering that the integration 
time of the TEAM1k chip is 14 times longer, the beneficial effect of cooling on the sensor 
radiation tolerance is evident. 

\begin{figure}[h]
\begin{center}
\epsfig{file=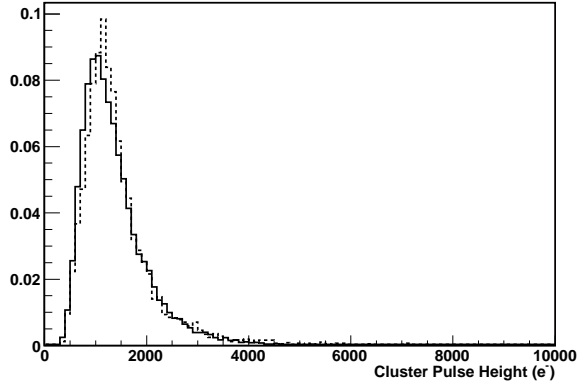, width=8.5cm}
\end{center}
\caption[]{Single electron cluster signal (in $e^-$ units) for 300~keV electrons before 
(continuous line) and after 5~Mrad of dose (dashed line) and recalibration. The comparison 
of the width of the two distribution shows that the cluster noise after irradiation is 
comparable to that of the unirradiated detector.}
\label{fig:ph_irradiation}
\end{figure}

From a device operation point of view, the increase of the leakage current results in 
an increase of the pixel base level. This is removed by subtraction of the pedestal level. 
However, the leakage current increase causes a decrease of the dynamic range and an increase 
of the single pixel noise. At 25~MHz clock frequency, after 5~Mrad of dose we observe a decrease
of the pixel dynamic range by about 30\%, which still allows proper operation of the chip.
\begin{table}[h]
\caption[]{Charge-to-voltage conversion gain and pixel noise as a function of the
delivered 300 keV electron dose.}
\begin{center}
\begin{tabular}{|c|c|c|}
\hline
\textbf{Dose}    & \textbf{Calibration}    & \textbf{Noise} \\
       (Mrad)     & ($e^-$/ADC)              & ($e^-$ ENC)     \\
\hline
0                 &      11.2       &         56$\pm$10    \\ 
1                 &      8.0        &         60$\pm$7     \\ 
2                 &      8.2        &         75$\pm$7     \\
3                 &      7.3        &         70$\pm$6     \\
4                 &      6.7        &         67$\pm$6     \\
5                 &      7.3        &         69$\pm$6     \\
\hline
\end{tabular}
\end{center}
\label{tab:noise}
\end{table}
Table~\ref{tab:noise} summarises the pixel charge-to-voltage gain calibration
and noise as a function of the dose, for a pixel clock frequency of
25~MHz. The gain calibration is estimated from the position of the most probable 
value of the Landau distribution for single electrons measured at low flux after 
each irradiation step. A slight decrease of the pixel gain is observed with the 
increasing dose, while the noise increases by about 25\%. The single electron  
cluster pulse height is not significantly affected after irradiation 
(see Figure~\ref{fig:ph_irradiation}) and the average cluster signal-to-noise 
ratio (S/N) measured for single 300~keV electron detection changes from 12.3 
before irradiation to 10.9 after 5~Mrad of dose. Finally, we check the 
imaging properties of the detector by determining the LSF after irradiation. We 
repeat the fit to the measured pulse height on pixel rows across an image of the 
stretched gold wire. By performing a single sigmoid fit, as discussed in section~5.1, 
we obtain a LSF value of (7.2$\pm$0.6)~$\mu$m at 300~keV, which is consistent with 
the value of (7.4$\pm$0.6)~$\mu$m obtained before irradiation.

\section{Conclusions}

A direct detection detector which meets the requirements for TEAM has been developed and 
characterised using 80~keV to 300~keV energy electrons. We measure a line spread function 
of (12.1$\pm$0.7)~$\mu$m to (7.4$\pm$0.6)~$\mu$m for 80 $\le E_e \le$ 300~keV with bright 
field illumination and  (6.7$\pm$0.3)~$\mu$m to (2.4$\pm$0.2)~$\mu$m with cluster imaging
and a DQE = 0.78$\pm$0.04 and 0.74$\pm$0.03 at the two ends of the energy range explored. 
The imaging performances of the detector are identical after irradiation with 300~keV electrons
up to a dose of 5~Mrad, while the dynamic range is reduced by $\simeq$~30~\% due to leakage
current increase.
The good agreement obtained in pixel response and line spread function  between measurements 
and simulation demonstrate that the tools to understand the detailed performance of the 
detector are in hand. R\&D on new detectors with enhanced performances is currently under way.

\section*{Acknowledgements}

\vspace*{-0.1cm}

We wish to thank N.~Andresen and R.~Erni who contributed to the detector installation on 
the electron microscope. We are grateful to the INFN group of Padova, Italy for their support 
with the DAQ system, in particular to D.~Bisello, D.~Pantano and M.~Tessaro. This work was 
supported by the Director, Office of Science, of the U.S. Department of Energy under Contract 
No.DE-AC02-05CH11231. The TEAM Project is supported by the Department of Energy, Office 
of Science, Basic Energy Sciences.

\vspace*{-0.1cm}

 \nolinenumbers


\begin{thebibliography}{99}

\bibitem{team}
  U.~Dahmen,  Microscopy and Microanalysis {\bf 13} (2) (2007).

\bibitem{team2}
  C.~Kisielowski, R.~Erni and B.~Freitag, Microscopy and Microanalysis {\bf 14} (2) 
  (2008) 78.

\bibitem{nature}
 V.E.~Cosslett {\it et al.}, Nature {\bf 281} (1979) 49.

\bibitem{science}
 C.O.~Girit {\it et al.}, Science {\bf 323} (2009) 1705.

\bibitem{fossum}
  E.R.~Fossum, IEEE Trans.\ Electron. Devices {\bf 44} (1997) 1689.

\bibitem{cmos}
  R.~Turchetta {\it et al.}, 
  Nucl.\ Instrum.\ and Meth.\ A {\bf 458} (2001) 677.

\bibitem{emicro}
  A.-C.~Milazzo {\it et al.}, Ultramicroscopy {\bf 104} (2005) 152. 
 
\bibitem{deptuch}
  G.~Deptuch {\it et al.}, Ultramicroscopy {\bf 107} (2007) 674. 

\bibitem{Denes:2007}
  P.~Denes, J.-M.~Bussat, Z.~Lee, V.~Radmilovic, 
  Nucl.\ Instrum.\ Meth.\ A {\bf 579} (2007) 891.

\bibitem{fan}
  G.Y.~Fan {\it et al.}, Ultramicroscopy {\bf 70} (1998) 113. 

\bibitem{faruqi}
  A.R.~Faruqi, D.M. Cattermole, C.~Reburn, 
  Nucl.\ Instrum.\ Meth.\ A {\bf 513} (2003) 317.

\bibitem{faruqi2}
  A.R.Faruqi {\it et al.}, Ultramicroscopy {\bf 94} (2003) 263. 
 
\bibitem{review}
  A.R.~Faruqi and R.~Henderson, Current Opinion in Structural Biology 
  {\bf 17} (2007) 549. 

\bibitem{Battaglia:2008yt}
  M.~Battaglia {\it et al.},
  Nucl.\ Instr.\ Meth.\  A {\bf 598} (2009) 642
  [arXiv:0811.2833 [physics.ins-det]].

\bibitem{Battaglia:2009aa}
  M.~Battaglia {\it et al.},
  Nucl.\ Instr.\ Meth.\  A {\bf 605} (2009) 350
  [arXiv:0904.0552 [physics.ins-det]].

\bibitem{Battaglia:2006tf}
  M.~Battaglia {\it et al.},
  Nucl.\ Instrum.\ Meth.\  A {\bf 579} (2007) 675
  [arXiv:physics/0611081].

\bibitem{aptek-ref}
  Aptek Industries, San Jose, CA 95111, USA.

\bibitem{Agostinelli:2002hh}
  S.~Agostinelli {\it et al.},
  Nucl.\ Instrum.\ Meth.\ A {\bf 506} (2003) 250.

\bibitem{Chauvie:2001fh}
  S.~Chauvie {\it et al.},
  {\it Prepared for CHEP'01: Computing in High-Energy Physics and Nuclear, Beijing, China, 3-7 Sep 2001}

\bibitem{Battaglia:2007eu}
  M.~Battaglia,
  Nucl.\ Instrum.\ Meth.\  A {\bf 572} (2007) 274.

\bibitem{Gaede:2006pj}
  F.~Gaede,
  Nucl.\ Instrum.\ Meth.\ A {\bf 559} (2006) 177.

\bibitem{Battaglia:2009zz}
  M.~Battaglia {\it et al.},
  Nucl.\ Instrum.\ Meth.\  A {\bf 611} (2009) 105.

\bibitem{avnet}
  manufactured by Avnet Inc., Phoenix, Arizona 85034 USA.
 
\bibitem{Brun:1997pa}
  R.~Brun and F.~Rademakers,
  Nucl.\ Instrum.\ Meth.\  A {\bf 389} (1997) 81.

\bibitem{Gaede:2005zz}
  F.~Gaede {\it et al.},
  in Proc. of {\it  Interlaken 2004, Computing in high energy physics and nuclear physics}, 
  Report CERN 2005-002, 471.

\bibitem{Battaglia:2009dt}
  M.~Battaglia, D.~Contarato, P.~Denes and P.~Giubilato,
  Nucl.\ Instrum.\ Meth.\  A {\bf 608} (2009) 363
  [arXiv:0907.3809 [physics.ins-det]].

\bibitem{optik}
  A.L.~Weickenmeier, W.~N\"uchter, J.~Mayer, Optik {\bf 99} (4) (1995) 147.

\bibitem{dqe}
  K.-H.~Herrmann and D.~Krahl, Adv.\ Opt.\ Electr.\ Microsc.\ {\bf 9} (1984) 1.

\bibitem{ishizuka}
  K.~Ishizuka, Ultramicroscopy {\bf 52} (1993) 7.

\bibitem{horacek}
  M.~Horacek, Rev.\ Sci. Instr.\ {\bf 76} (2005) 093704.

\end{thebibliography}
\end{document}